\documentclass[prl,twocolumn,showpacs,groupedaddress]{revtex4}

\usepackage{graphicx}

\begin{document}

\bibliographystyle{apsrev}

\title{Dynamics of the Destruction and Rebuilding of a Dipole Gap in Glasses}

\author{S. Ludwig, P. Nalbach, D. Rosenberg, D. Osheroff}
\affiliation{Departement of Physics, Stanford University, Stanford, California
  94305-4060, USA}

\date{\today}

\begin{abstract}
After a strong electric bias field was applied to a glass sample at
temperatures in the millikelvin range its AC-dielectric constant increases and then decays logarithmically with time.
For the polyester glass mylar we have observed the relaxation of the dielectric constant
back to its initial value for several temperatures and 
histories of the bias field. Starting from the dipole gap theory we have developed a model  
suggesting that the change of the dielectric constant after transient
application of a bias field is only partly due to relaxational  
processes. In addition, non-adiabatic driving of tunneling states (TSs) by applied electric fields causes long
lasting changes in the dielectric constant. 
Moreover, our observations indicate that at temperatures below $50$~mK the
relaxation of TSs is caused primarily by interactions between TSs.

\end{abstract}


\maketitle

\def\kb{k_{\rm B}}
\def\tw{t_{\rm w}}
\def\ts{t_{\rm s}}
\def\Deltac{\Delta_{0\rm c}}
\def\de{\delta\epsilon / \epsilon}

At low temperatures the internal energy of glasses is vastly elevated by atomic tunneling states (TSs) and most properties of
amorphous solids are strongly influenced by these additional degrees of freedom~\cite{Zel71,Hun00}. The phenomenological 'tunneling 
model' describes the thermal, elastic and dielectric properties of glasses at low temperatures successfully~\cite{Phi72,And72}. 
Starting from the central assumption that the potential minima of groups of atoms are not well defined, it describes such a
configuration by a 'particle' moving in a double well potential. The two identical and harmonic wells may differ in depth by
the asymmetry energy $\Delta$. At low temperatures only the energy ground states of the two wells are occupied and the TSs
oscillate by quantum mechanical tunneling between them. The energy splitting of these two-level systems is given by 
$E_0=\sqrt{\Delta_0^2+\Delta^2}$, where $\Delta_0$ is the tunneling splitting. The parameters $\Delta_0$ and $\Delta$ are widely
distributed due to the randomness of the glassy structure. 

Soon after the introduction of the tunneling model in 1972 it was shown 
that the mutual interaction of TSs causes spectral diffusion, 
while interaction that involves energy exchange between TSs was long believed to be unimportant.
However, more recent experiments revealed deviations from the tunneling model and indicate that for many properties the 
interaction between TSs gains importance with decreasing temperature~\cite{Ens02,OshDip}. 

Here we will concentrate on the effects of electric bias fields on the low temperature AC-dielectric constant of amorphous 
materials. A sudden large change of the electric bias field causes the 
relative dielectric constant $\de$
of a glass sample, that was originally in thermal equilibrium, to immediately change to a larger value and then decay 
back to its initial value logarithmically for several decades in time. In addition, a relatively slow sweep of the bias field 
reveals a minimum of $\de$ at the initial bias field, in our case at zero field~\cite{OshDip}. A.
Burin explained these findings within a 'dipole gap' model, which considers the weak dipole-dipole interaction 
$J_{\rm ij}=U_0/r_{\rm ij}^3$ between TSs~\cite{Bur95}. Although the interaction is weak, some TSs fulfill the condition 
$J_{\rm ij}\gtrsim (E_{0\rm i}+E_{0\rm j})$ for strongly coupled pairs. Within the dipole gap model, such strongly coupled
TSs form clusters that do not contribute fully to the dielectric constant because their individual dipole moments are 
pinned relative to each other. A change of the bias field alters the distribution of asymmetry energies and leads 
to the formation of new clusters of strongly coupled TSs. 
As long as the new clusters remain out of thermal equilibrium, their
TSs contribute fully to the dielectric constant~\cite{Bur95}. This explains the observed minimum of $\de$ at the
initial bias field. In addition mutual interaction of TSs leads to a new relaxation channel~\cite{Bur94} due to resonant energy exchange
between pairs of TSs resulting in the overall relaxation rate
\begin{equation}
\tau^{-1} = \gamma_{0} \Delta_0^2 E \, \coth\left( {E\over{2\kb T}}\right) \, + \, \alpha_0 {\Delta_0^2\over E^2} \kb T \; , 
\label{eq1}
\end{equation}
consisting of the well known one-phonon relaxation rate (first term r.h.s.) and the contribution due to interaction. The  
material constants $\gamma_{0}$~\cite{Jac72} and $\alpha_{0}$~\cite{Bur94} depend on the coupling of the TSs to phonons and 
between each other respectively.  Note that the dynamics of TSs that generate any change in the dielectric 
constant is dominated by thermally active TSs that fulfill $E=\sqrt{\Delta_0^2+(\Delta+{\bf p}\cdot{\bf F})^2}\sim\kb T$ producing an additional temperature 
dependence in~(\ref{eq1}) implicit in the energy splittings $E$. 
Moreover, after changing the bias field {\bf F} by more then ${\bf p}\cdot\delta{\bf F} \sim \kb T$, where ${\bf p}$ is the dipole
moment of the TSs, a different set of TSs fulfills $E\sim\kb T$. 
Therefore in past experiments in which a large bias field was suddenly applied to study the dynamics of strongly coupled  
clusters the formation of a new dipole gap consisting of different clusters than those formed at zero field was observed. The 
interpretations of these measurements relied on assumptions such as a flat energy distribution of TSs for $E_0\lesssim {\bf 
p}\cdot {\bf F}\equiv \Phi$.

All our measurements were made at zero bias field. Thus we always measured the same set of TSs.
However, we applied a large electric bias field $\Phi\gg\kb T$ for 
the waiting time $\tw$ to destroy the zero field dipole gap and then
observed the restoration of the same dipole gap 
by measuring $\delta\epsilon / \epsilon$ in zero field. In addition, we took
care to linearly sweep the bias field up and down in the same sweep time $\ts$. 

As long as the bias field is applied the TSs are relaxing towards a new
equilibrium, thereby breaking clusters of TSs that were initially strongly coupled. Back at zero bias field these clusters 
rearrange. The TS's relaxation times depend according to~(\ref{eq1}) on their asymmetry energies and are broadly distributed. 
For $\Phi\gg\kb T$ initially thermal TSs with $E_0\sim\kb T$ have $E\sim\Phi$ at the bias field {\bf F}.
We introduce the decay time $\tau_0$ at which we expect the dielectric constant to again reach its initial value, i.e. 
$\epsilon(t)=\epsilon(\infty)$ before the bias field was applied 
\begin{equation}\label{eq2} 
\frac{\tau_0}{\tw} = \frac{\tau(E=\kb T)}{\tau(E=\Phi)} \simeq
\left(\frac{\kb T}{\Phi}\right)^2 
\frac{(\gamma_0\Phi^3+\alpha_0\kb T)}{2.2\gamma_0(\kb T)^3+\alpha_0\kb T} \; .
\end{equation}
For this we assumed that both the destruction and restoration of the dipole gap are solely caused by relaxational
processes, thus that $\tau_0$ is given by the ratio of the relaxation times of
TSs without and with field times the waiting time.
Following the dipole gap theory we obtain for the change of the dielectric
constant, after a bias field {\bf F} was applied for a time $\tw$,
\begin{equation} 
\de \,\simeq\,  2P_0 \frac{2p^2}{3\varepsilon_0\varepsilon}   
\frac{P_0U_0}{2} \frac{2\pi}{3} \left[ \ln\left\{ \frac{\Phi}{\kb T} \right\}
\right]^2 \ln\left\{ \frac{\tau_0}{t} \right\} \; .
\label{eq3}
\end{equation}
Since in (\ref{eq3}) relaxational processes are approximated by Heaviside step
functions we expect deviations from logarithmic decays at the start
and the end of the decay. 

Our experiments were performed on $15 \mu$m thick mylar film (amorphous polyester) at a frequency of 10~kHz at temperatures well 
below the minimum of $\de$ at $T_{\rm min}(10 {\rm kHz})\simeq 50$~mK, ensuring that $\de$ consists entirely of its resonant part.
Gold capacitor electrodes were evaporated onto the sample. The dielectric constant was
measured with a hand made bridge by comparison to 
a vacuum reference capacitor, that was placed on the one-kelvin pot of the dilution refrigerator. 
The AC-field was always kept within its linear response region. 
%
\begin{figure}[th]
\includegraphics[width=0.95\linewidth]{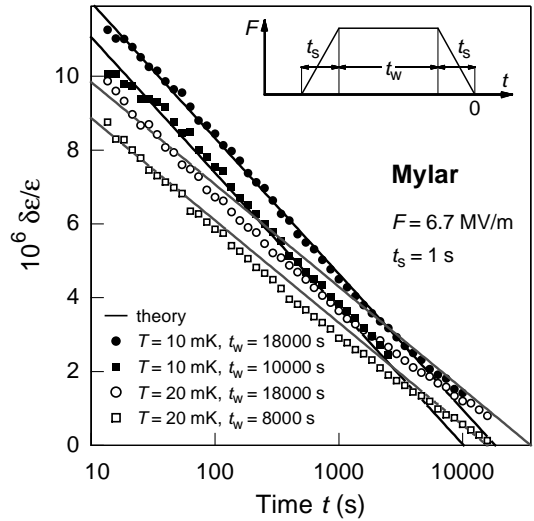}
\vskip -8mm \caption{Relative dielectric constant of Mylar after a bias field $F$ was applied at two temperatures 
for very long waiting times $\tw\gg\ts$. All lines are theory curves
after~(\ref{eq3}) (see text). The shematic indicates the application of the bias field and the choice of $t=0$.}
\label{fig1}
\end{figure}
%

Fig.~\ref{fig1} shows the evolution of $\de$ after a bias field of $F=6.7$~MV/m was applied for two different waiting times of
several hours ($\tw\gg\ts=1$~s) at each of the temperatures $T=10$~mK and $T=20$~mK. 
The zero point of time ($t=0$) is chosen to be at the completion of the bias field sweep and $\de=0$ is 
defined as its initial value at zero field. The amplitude of $\de\sim 10^{-5}$ increases for longer waiting times and decreasing temperature as 
expected from~(\ref{eq3}). The dielectric constant decays approximately logarithmically over several decades in time justifying the 
simplification of describing relaxational processes by Heavyside Step Functions. Taking only the one-phonon process 
into account ($\alpha_0=0$) we would expect decay times of $\tau_0\simeq 73\tw$ for $T=10$~mK and $\tau_0\simeq 37\tw$ for
$T=20$~mK according to~(\ref{eq2}). In contrast, our measurements suggest decay times in the order $\tau_0\simeq\tw$ indicating the
importance of interaction mediated relaxation~\cite{footnote1}. 
Furthermore our experiments indicate that the restoration of the dipole gap takes longer for the higher temperature. 
According to~(\ref{eq1}) it is the interaction mediated relaxation process that results in an increasing relaxation
time for the contributing TSs ($E_0\sim\kb T$) with increasing temperature. 
Therefore our data unambiguously show that interaction mediated relaxation dominates in this temperature range.

One of the solid lines in Fig.~\ref{fig1} is a linear fit for $T=10$~mK and $\tw=18000$~s (filled circles). 
By comparison with~(\ref{eq3}) we find
the fit-parameters $\tau_0/\tw \simeq 1.02$ and $P_0 U_0 \simeq 4.1\cdot10^{-4}$, the latter lying whithin typical estimations for 
other glasses\cite{Bur95}.
The prefactor in~(\ref{eq3}) was previously obtained from the temperature variation of $\de$ well below $T_{\rm min}$ as 
$2P_0 \frac{2p^2}{3\varepsilon_0\varepsilon} \simeq 1.35\cdot10^{-4}$~\cite{OshDip,footnote2} and the dipole moment of the TSs in
mylar was estimated to be $p\simeq 1.2$~Debye as discussed later (see fig.~\ref{fig4}).  
Using~(\ref{eq1}) we estimate the temperature at which the interaction mediated relaxation time equals the one-phonon relaxation 
time as 55~mK~$\lesssim T_{\rm c}\lesssim 95$~mK for $3\gtrsim\tau_0/\tw\gtrsim1$. Below $T_{\rm c}$ the interaction mediated 
relaxation dominates.
The other three lines in fig.~\ref{fig1} are calculated from~(\ref{eq3}) using the same set of parameters. Our simplifying model agrees remarkably 
well with the data besides deviations from logarithmic decays particularly at the higher temperature. 

\begin{figure}[th]
\includegraphics[width=0.95\linewidth]{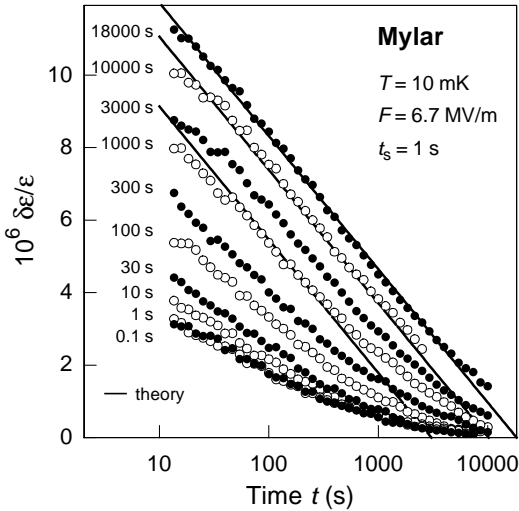}
\vskip -8mm \caption{Relative dielectric constant of Mylar after a bias field $F$ was applied for waiting times
$\tw$ as marked left of each curve, while $\ts=1$~s for all data. The solid lines are theory curves
after~(\ref{eq3}) using the same parameters as in fig.~\ref{fig1}.}
\label{fig2}
\end{figure}

\begin{figure}[th]
\includegraphics[width=0.95\linewidth]{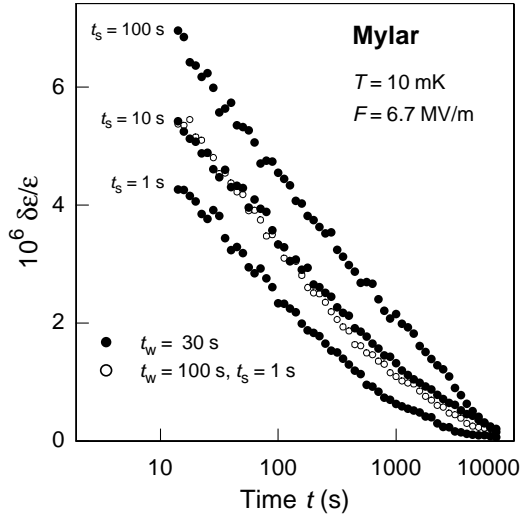}
\vskip -8mm \caption{Relative dielectric constant of Mylar after a bias field $F$ was applied for different $\ts$ and $\tw$.}
\label{fig3}
\end{figure}

\begin{figure}[th]
\includegraphics[width=0.95\linewidth]{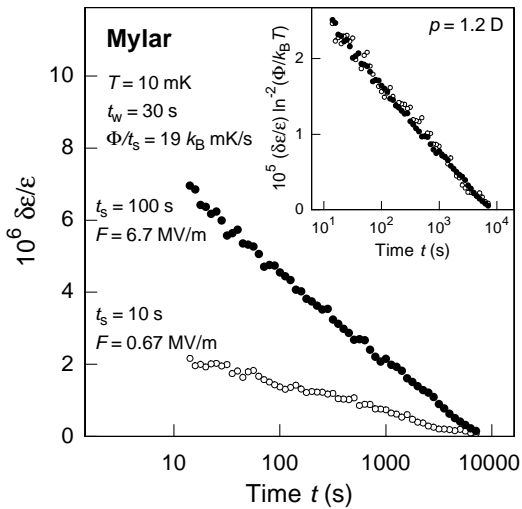}
\vskip -8mm \caption{Relative dielectric constant of Mylar after a bias field
  $F$ was applied for the same sweep  
rate, the same waiting time but different bias field strengths and sweep
times. In the inset the data are normalized with respect to 
their bias field for the dipole moment of $p=1.2$~D.}
\label{fig4}
\end{figure}

In fig.~\ref{fig2} the restoration of the dipole  
%
%
gap at the temperature of $T=10$~mK is displayed for 10 different waiting times between $0.1$~s and 5 hours and otherwise the same conditions as for the data in fig.~\ref{fig1}. 
The straight lines in fig.~\ref{fig2} are calculated from~(\ref{eq3}) for the three longest waiting times using the 
same parameters as for the theory curves in fig.~\ref{fig1}. 
Our model which considers relaxation predicts parallel decay curves shifted along the logarithmic time axis proportional to $\ln(\tw)$.
In experiments we observe this behaviour only for our longest waiting times (fig.~\ref{fig1}). 
As $\tw$ gets shorter $\de$ increasingly decays slower than expected, until for $\tw\lesssim\ts$ (our two shortest waiting times in 
fig.~\ref{fig2}) the waiting time dependence disappears completely.  
To further investigate this puzzling effect we show in Fig.~\ref{fig3} the decay curves for 3 different sweep times at
$\tw=30$~s (solid points). 
As with longer $\tw$, $\de$ increases for longer $\ts$. For comparison we included the data points from Fig.~\ref{fig2}  
for $\ts=1$~s and $\tw=100$~s (open circles). Surprisingly their course is almost identical to the
curve for $\ts=10$~s and $\tw=30$~s with a much smaller waiting time even if one were to make the obvious overestimation 
$\tw+2\ts\rightarrow\tw$ to include relaxation while the bias field is being changed. 

At this point we can safely conclude that at waiting times shorter than $\tw\sim 3000$~s an additional
process gains importance that is not relaxational (due to waiting time
independence) and instead depends on the sweep rate of the bias field.
We propose that the changing electric (AC and DC) field drives TSs non-adiabatically during the bias field sweep, and that this
process is responsible for the new behavior.
%
%

Let us consider the time dependent quantum mechanics of the TSs while the bias field is being changed.  
If the field sweep is 'slow' the TSs are driven adiabatically and their wave functions always adjust to the momentary field. 
In the opposite case of a 'fast' sweep the TSs are driven non-adiabatically and their wave functions stay unchanged. These TSs 
remain in their original quantum states.
Most TSs are asymmetric and therefore localized. In the following we consider non-adiabatically driven TSs having  
asymmetry energies that decrease to zero and successively increase again with changing field.  
For these TSs the original ground (excited) state at zero field converts
into the excited (ground) state at maximum field. As the result the occupation numbers of these TSs are inverted. 
However, as the field is changed 
back to zero the non-adiabatically driven TSs are in their original state having been inverted twice. 
The Landau-Zenner criterion~\cite{LanZen} gives the probability
\begin{equation} R_1(\Delta_0) \,=\, e^{-\left(
      \frac{\Delta_0}{\Deltac} \right)^2}\qquad {\rm with}\quad \Deltac=\pi\sqrt{\Phi h/\ts}\;
\label{eq4}
\end{equation}
with Plank's constant $h$ that a system is driven non-adiabatically near $\Delta+\Phi=0$. Only TSs with
$\Delta_0\sim\Deltac$ have a finite probability to have been inverted exactly once~\cite{footnote3}. 
The non-adiabatically driven TSs have very long
relaxation times $\tau(\Deltac,E_0)$ which are according to 
(\ref{eq1}) and (\ref{eq4}) proportional to $\tau\sim\ts/\Phi$. 
Neglecting the relaxational process leading to (\ref{eq2}) we expect the
decay time at which the dielectric constant reaches its initial value to be a
function of $\ts/\Phi$ as well. However, the predicted decay would be
non-logarithmic resembling a smoothened step occuring at the time
$t\sim\tau(\Deltac,E_0=k_{\rm B}T)$. 

An AC measuring field forces the dipole moment of a TS to oscillate with the
measuring frequency $\omega$ as long as the quantum mechanical time evolution
of the TS is able to follow the field. However,
this description fails when the energy splitting $E$ of a TS is smaller than a
critical value $E_{\rm AC}=\sqrt{\hbar\omega\,{\bf p}\cdot{\bf F}_{\rm AC}}$ 
below which the AC-measuring field $F_{\rm AC}$ drives the system
non-adiabatically.  
TSs with $\Delta_0<E_{\rm AC}$ are driven
nonadiabatically by the AC field for the time
$\delta t^\star=\ts\,F_{\rm AC}/F_{\rm DC}$ in which $E<E_{\rm AC}$. An initially localised TS 
will oscillate between the two wells during $\delta t^\star$ with a renormalized
tunneling frequency $\widetilde{\Delta}_0/h$~\cite{Grossmann91}. Accordingly,
TSs with $h\widetilde{\Delta}^{-1}_0>\delta t^\star$ stay localised in the
original state whereas TSs with $h\widetilde{\Delta}^{-1}_0\ll\delta t^\star$
oscillate many times so that only half of them end up in their initial state. 
The AC-driving effect predicts a logarithmic decay with the decay time $\tau_0\equiv\tau_0(\delta
t^\star\sim\ts/\Phi)$~\cite{footnote4}.

A combination of DC- and AC-driving explains the sweep time dependence visible in fig.~\ref{fig2} and~\ref{fig3} qualitatively. 
In our experiments $F_{\rm AC}\simeq 9.7$~kV/m, resulting in $E_{\rm AC}\sim\Deltac$ for $F_{\rm DC}\simeq 6.7$~MV/m and $\ts=1$~s. 
At longer $\ts$ we expect AC-driving to increasingly become important at the expense of the DC-driving effect. 
The data in fig.~\ref{fig3} indicate a  
crossover from non-logarithmic to logarithmic behavior at $\ts\sim 10$~s. Consistently we did not observe  
a dependence on $F_{\rm AC}$ for $\ts\le 10$~s (unpublished data) but have not searched for that effect for $\ts>10$~s.  
%
%

Our model including non-adiabatic driving predicts $\tau_0\equiv\tau_0(\ts/\Phi_{\rm DC})$. 
In fig.~\ref{fig4} we plot two data sets with the same $\Phi/\ts=19~\kb$mK/s but different sweep times and bias fields. 
The decay times of both data curves are the same as expected. 
We stress that relaxational processes during $\tw$ cause a decay time $\tau_0\sim\tw$
(fig.~\ref{fig1}), but in fig.~\ref{fig4} $\tau_0\gg\tw$ and thus the decay is due to the driving effect.
In our model this can be understood by recognizing that the finite waiting time necessarily implies a smallest
$\Delta_0$ for TSs that can possibly decay while the bias field is applied. This smallest $\Delta_0$ is larger than $\Deltac$
unless $\tw\gg\ts$. 
Regardless of which effect breaks the strongly coupled clusters of TSs, we always find the temperature dependence of the decay of 
the dielectric constant propotional to $[\ln(\Phi/\kb T)]^2$. This is a consequence of the central assumption in the dipole gap
model that only TSs with energy splitting $T\lesssim E_0\lesssim\Phi$ contribute to changes in the dielectric constant~\cite{Bur94}.  
Using this dependence we normalized the two data sets as shown in the inset in fig.~\ref{fig4}. Very good agreement was reached 
with the dipole moment $p\simeq 1.2$~Debye (that was used for the theory curves in fig.~\ref{fig1} and~\ref{fig2}). 

We have investigated the dynamics of the destruction and the restoration of the
zero field dipole gap by application of strong electric  
bias fields. Only our data curves at very long waiting times $\tw\gg\ts$ can be
explained by solely relaxational processes.  
Our investigation suggests non-adiabatic driving of the TSs as a possible
reason for the destruction of the dipole gap for  
shorter $\tw$. Furthermore we have shown that at temperatures below $50$~mK the
relaxation of TSs is dominately by a relaxation mechanism
due to interactions between TSs~\cite{Bur94}.

This work was supported by U$.$S$.$ Dept$.$ of Energy grant DE-FG03-90ER45435-M012. S$.$ Ludwig
thanks the Deutsche Forschungsgemeinschaft and P$.$ Nalbach the Alexander von
Humboldt Foundation for support.


\begin{thebibliography}{99}
\bibitem{footnote1}{The first data taken after cooling are never displayed in
        our figures.  
        They are shifted to higher amplitudes of $\de$ than all
        successive measurements, which were  
        performed with sufficient time in between to assure
        always the same 
        starting conditions. The first data after cooling yield an about
	a factor of three longer $\tau_0$.}
\bibitem{footnote2}{The value listed in~\cite{OshDip} is too large by $\ln(10)$.}
\bibitem{footnote3}{Due to spectral diffusion coherence times of relevant TSs
        are short enough that  
        quantum mechanical interference between the flips while ramping the
        bias field up versus down can be neglected.}
\bibitem{footnote4}{Details will be published elsewhere by the authors.}
\end{thebibliography}
\end{document}